\documentstyle[preprint,aps,eqsecnum,epsfig]{revtex}
\tightenlines
\begin{document}
\draft 
\title{\bf NON-EQUILIBRIUM PHASE TRANSITIONS IN CONDENSED MATTER AND COSMOLOGY:
SPINODAL DECOMPOSITION, CONDENSATES AND DEFECTS\footnote{Lectures
delivered at the NATO Advanced Study Institute:\\ 
Topological Defects and the Non-Equilibrium Dynamics of Symmetry
Breaking Phase Transitions} } 
\author{{\bf D. Boyanovsky$^{(a,b)}$, H.J. de Vega$^{(b,a)}$ and
R. Holman$^{(c)}$ }}  
\address { (a)Department of Physics and Astronomy,
University of Pittsburgh, Pittsburgh, PA 15260 USA\\ (b)
LPTHE\footnote{Laboratoire Associ\'{e} au CNRS UMR 7589.}  Universit\'e Pierre
et Marie Curie (Paris VI) et Denis Diderot (Paris VII), Tour 16, 1er. \'etage,
4, Place Jussieu 75252 Paris, Cedex 05, France \\ 
(c) Department of Physics,
Carnegie Mellon University, Pittsburgh, PA 15213, USA} 
\date{\today} 
\maketitle
\begin{abstract}
These lectures address the dynamics of phase ordering out of equilibrium in condensed matter
and in quantum field theory in cosmological settings, emphasizing their similarities and 
differences. In  condensed matter we describe the phenomenological approach based on the
Time Dependent Ginzburg-Landau (TDGL) description. After a general discussion of the main 
experimental and theoretical features of phase ordering kinetics and the description of 
linear (spinodal) instabilities we introduce the scaling hypothesis and show how a dynamical
correlation length emerges in the large $ N $ limit in  condensed matter systems. The large $N$
approximation is a powerful tool in quantum field theory that allows the study of non-perturbative 
phenomena in a consistent manner. We study the  {\em exact} solution to the dynamics after a quench 
in this limit  in Minkowski  space time and  in radiation dominated Friedman-Robertson-Walker 
Cosmology.  There are some remarkable similarities between these very different settings such as the
 emergence of a scaling regime and of  a dynamical correlation length at late times that describe the 
formation and growth of ordered regions. In quantum field theory and cosmology  this length scale is 
constrained by causality and its growth in time is also associated with coarsening and the onset of a 
condensate. We provide a density matrix interpretation of the formation of defects and the 
classicalization of quantum fluctuations. 
\end{abstract}

\section{Phase Ordering Kinetics: an interdisciplinary fascinating problem}
The dynamics of non-equilibrium phase transitions and the ordering
process that occurs until the system reaches a 
broken symmetry equilibrium state play an important role in many
different areas. Obviously in the physics of binary 
fluids and ferromagnets (domain walls) 
superfluids (vortex formation), and liquid crystals (many possible
defects)  to name but a few in condensed matter, but also 
in cosmology and particle physics. In cosmology defects produced 
during Grand Unified Theory (GUT) or the Electro-weak 
(EW) phase transition
can act as seeds for the formation  of large scale structure 
and the dynamics of phase ordering and formation of
ordered regions is at the heart of Kibble's mechanism of 
defect formation\cite{kibble,kibble2,kibble3,vilen}. Current and
future measurements of Cosmic Microwave Background 
anisotropies will  determine the nature of the  cosmological phase
transitions that influenced structure formation\cite{durrer}.  
 Also at even lower energies, available with current
and forthcoming accelerators (RHIC and LHC) the phase transitions
predicted by the theory of strong interactions, Quantum Chromodynamics
(QCD) could occur out of equilibrium via the formation of coherent
condensates of low energy Pions. These conjectured configurations
known as `Disoriented Chiral Condensates'
are similar to the defects expected in liquid crystals or ferromagnets
and their charge distribution could be an  
experimental telltale of the QCD phase
transitions\cite{rajagopal}. Whereas the GUT phase transition took
place when the Universe was 
about $10^{-35}$ seconds old and the temperature about $10^{23}K$, and
the EW phase transition occured when the 
Universe was $10^{-12}$ seconds old and with a temperature $10^{15}K$,
the QCD phase transition took place at about 
$10^{-5}$ seconds after the Big Bang, when the temperature was a mere
$10^{12}K $. This temperature range will be probed 
at RHIC and LHC within the next very few years. The basic problem of
describing the process of phase ordering and the 
competition between different broken symmetry states on the way towards  
reaching equilibrium is common to all of these situations. The tools,
however, are necessarily very different: whereas 
ferromagnets, binary fluids or alloys etc, can be described via a
phenomenological (stochastic) description, certainly in quantum 
systems a microscopic formulation must be provided. In these lectures
we describe  a program to include 
 ideas from condensed matter to the realm of quantum field theory, to
describe phenomena on a range of time and spatial 
scales of unprecedented resolution (time scales $\leq  10^{-23}$
seconds, spatial scales $\leq 10^{-15}$ meters) that 
require a full quantum field theoretical  description. 

\section{Main ideas from Condensed Matter:}
Before tackling the problem of describing phase ordering kinetics in
quantum systems starting from a microscopic 
theory, it proves illuminating to understand a large
body of theoretical and experimental work in 
condensed matter systems\cite{bray}-\cite{marco}. Although ultimately
the tools to study the quantum problem will be  different, the 
main physical features to describe are basically the same: the
formation and evolution of correlated regions separated 
by `walls' or other structures. Inside these regions an ordered phase
exists which eventually grows to 
become macroscopic in size. Before attempting to describe the manner
in which a given system orders after being cooled 
through a phase transition an understanding of the relevant time
scales is required. Two important time scales determine 
if the transition occurs in or out of equilibrium: the relaxation time
of long wavelength fluctuations (since these 
are the ones that order) $\tau_{rel}(k)$ and the inverse of the
cooling rate $t_{cool}= T(t)/\dot{T}(t)$. 
If $\tau_{rel}(k)<<t_{cool}$ then these wavelengths are in local
thermodynamical equilibrium (LTE), but if  
$ \tau_{rel}(k)>>t_{cool} $ these wavelengths fall out of LTE and freeze out, for
these the phase transition occurs 
in a quenched manner. These modes do not have time to adjust
locally to the temperature change and for them the 
transition from a high temperature phase to a low temperature one
occur instantaneously. This description was 
presented by Zurek\cite{zurek}  analysing   the emergence of defect
networks after a quenched phase transition.   
Whereas the
short wavelength modes are rapidly thermalized (typically by
collisions) the long-wavelength modes with $ k << 1/\xi(T) $ with
$ \xi(T) $ the correlation length (in the disordered phase) become {\em
critically slowed down}. As $ T\rightarrow T_c^+ $ the long 
wavelength modes relax very slowly, they fall out of LTE and any
finite cooling rate causes them to undergo a quenched non-equilibrium
phase transition. As the system is quenched from $ T>T_c $ (ordered 
phase) to $ T<<T_c $ (disordered phase) ordering {\em does not} occur
instantaneously. The length scale of the ordered 
regions grows in time (after some initial transients) as the different
broken symmetry phases compete to select the 
final equilibrium state. A {\em dynamical} length scale $ \xi(t) $ typically
emerges  which is interpreted as the size of 
the correlated regions, this dynamical correlation length grows in
time to become macroscopically large\cite{bray,langer,mazenko,marco}.  

The phenomenological  description of phase ordering kinetics begins
with a coarse grained local free energy functional of 
a (coarse grained) local order parameter $M(\vec r)$\cite{bray,langer}
which determines the {\em equilibrium} states. In Ising-like systems
this $M(\vec r)$ is the local magnetization 
(averaged over many lattice sites), in binary fluids or alloys
it is the local concentration difference, in superconductors is the
local gap, in superfluids is the condensate fraction etc. The typical free energy is (phenomenologically) of the Landau-Ginzburg form:
\begin{eqnarray}
F[M] & = & \int d^d\vec x \left\{\frac{1}{2} [\nabla M(\vec x)]^2 +
V[M(\vec x)]\right\} \nonumber \\ 
V[M] & = & \frac{1}{2} \;  r(T)\,  M^2 + \frac{\lambda}{4}\,M^4  ~~; ~~ r(T) =
r_0(T-T_c) \label{freenergy} 
\end{eqnarray} 

The equilibrium states for $T<T_c$ correspond to the broken symmetry states with $M= \pm M_0(T)$ with

\begin{equation}
M_0(T) = \left\{ \begin{array}{cc}
0 & ~\mbox{for} ~ T>T_c \\
\sqrt{\frac{r_0}{\lambda}}(T_c-T)^{\frac{1}{2}} &  ~\mbox{for} ~T<T_c
\end{array} 
\right. \label{MofT}
\end{equation}
 
Below the critical temperature the potential $V[M]$ features a non-convex
 region with $\partial^2 V[M] / \partial M^2 <0$  for
\begin{equation}
-M_s(T)<M<M_s(T) ~~; ~~ M_s(T) = 
\sqrt{\frac{r_0}{3\lambda}}(T-T_c)^{\frac{1}{2}} ~~~ (T<T_c)
\label{spinoregion}
\end{equation}
\noindent this region is called the spinodal region and corresponds to 
thermodynamically unstable states. The lines   $ M_s(T) $ vs. $ T $ and
$ M_0(T) $ vs. $ T $ [see eq.(\ref{MofT})] are known 
as the classical spinodal and coexistence  lines respectively.

 The states
between the spinodal and coexistence lines are metastable (in mean-field 
theory). As the system is cooled below $T_c$ into the unstable region inside
the spinodal, the {\em equilibrium} state of the system is a coexistence of
phases separated by domains and the concentration of phases is determined by the Maxwell 
construction and the lever rule. 




{\bf Question:} How to describe the {\em dynamics} of the phase transition and
the process of phase separation?

{\bf Answer:} A phenomenological but experimentally succesful description,
Time Dependent Ginzburg-Landau theory (TDGL) where the basic ingredient is
 Langevin dynamics\cite{bray}-\cite{marco}
\begin{equation}
\frac{\partial M(\vec r,t)}{\partial t} = -\Gamma[\vec r,M]  \; 
\frac{\delta F[M]}{\delta M(\vec r,t)} + \eta(\vec r,t) \label{langevin}
\end{equation}
with $ \eta(\vec r,t) $
a stochastic noise term, which is typically assumed to be white (uncorrelated)
and Gaussian and obeying the fluctuation-dissipation theorem:
\begin{equation}
\langle \eta(\vec r,t) \eta(\vec r',t') \rangle = 2\,T\, \Gamma(\vec
r) \, \delta^3(\vec r - {\vec r}^{\prime})\delta(t-t') ~~ ; ~~ \langle
\eta(\vec 
r,t) \rangle =0 \label{noise}
\end{equation} 
 \noindent the averages $\langle \cdots \rangle$ are over the Gaussian 
distribution function of the noise. There are two important cases to 
distinguish: {\bf NCOP:} Non-conserved order parameter, with $\Gamma=\Gamma_0$
a constant independent of space, time and order parameter, and which can
be absorbed  in a rescaling of  time. {\bf COP:} Conserved
order parameter with 
$$
\Gamma[\vec r] = - \Gamma_0 \; \nabla^2_{\vec r}
$$ 
where $\Gamma_0$ could depend on the order parameter, but  here we
will restrict the discussion to the case where it is a constant.
 In this latter case the average over the noise of the Langevin equation
can be written as a conservation law 
\begin{eqnarray}
\frac{\partial M}{\partial t} & = &  -\nabla\cdot J + \eta \; \Rightarrow \;
\frac{\partial}{\partial t} \langle \int d^3\; \vec r M(\vec r,t) \rangle =0 
\nonumber \\
\vec J & = &  \vec{\nabla}_{\vec r}
\left[-\Gamma_0\frac{\delta F[M]}{\delta M}
\right] \equiv \vec{\nabla}_{\vec r}\mu \label{equcop}
\end{eqnarray} 
\noindent where $\mu$ is recognized as the chemical potential. Examples
of the NCOP are the magnetization in ferromagnets, the gap in superconductors
and the condensate density in superfluids (the total particle number
is conserved but not the condensate fraction), of the COP: the concentration
difference in binary fluids or alloys. For a quench from $T>T_c$ deep into the
low temperature phase $T\rightarrow 0$ the thermal fluctuations are suppressed
after the quench and the noise term is irrelevant. In this situation of
experimental relevance of a deep quench the dynamics is now described by
a deterministic equation of motion, 

 for {\bf  NCOP}:
\begin{equation}
\frac{\partial M}{\partial t} = -\Gamma_0 \; \frac{\delta F[M]}{\delta
M}\label{NCOPeqn} 
\end{equation}

\noindent for {\bf  COP}:
\begin{equation}
\frac{\partial M}{\partial t} = \nabla^2\left[
\Gamma_0 \frac{\delta F[M]}{\delta M}\right] \label{cahnhil}
\end{equation}
\noindent which is known as the Cahn-Hilliard equation\cite{bray,langer}. In both cases
the equations of motion are purely diffusive
\begin{equation}
\frac{dF}{dt} = \int d^3 r \; \frac{\delta F[M]}{\delta M(\vec r,t)}  \;
\frac{\partial M(\vec r,t)}{\partial t} = -\Gamma_0
 \left\{ \begin{array}{cc}
& \int d^3 r \left(\frac{\delta F}{\delta M}\right)^2 
 ~\mbox{NCOP}  \\
& \int d^3 r \left(\vec{\nabla} \frac{\delta F}{\delta M}\right)^2 
 ~\mbox{COP} 
\end{array} 
\right. \label{diffu}
\end{equation}
\noindent and in both cases $\frac{dF}{dt}<0$. Thus, the energy is always
diminishing and there is no possibility of increasing the free energy. Thus
overbarrier thermal activation cannot be described in the absence of thermal
noise, which is clear since thermal activation is mediated by large thermal
fluctuations.  The fact that this phenomenological description is purely
dissipative with an ever diminishing free energy is one of the fundamental
differences with the quantum field theory description studied in the next
sections. 

\subsection{Critical slowing down in NCOP:}
\noindent Critical slowing down of long-wavelength fluctuations is
built in the TDGL description. Consider 
the case of NCOP and linearize the TDGL equation above the critical
temperature for small amplitude 
fluctuations near $M = 0$. Neglecting the noise term for the moment
and taking the Fourier transform of the small 
amplitude fluctuations we find
\begin{equation}
\frac{d m_k(t)}{dt} \approx -\Gamma_0\left[k^2+ r_0(T-T_c)\right]m_k(t) 
\end{equation}
showing that long-wavelength small amplitud fluctuations relax to
equilibrium $m_k =0$ on a time scale given by 
\begin{equation}
\tau_k \propto \left[k^2+ r_0(T-T_c)\right]^{-1} \label{relax}
\end{equation}
As $T \rightarrow T_c^+$ the long-wavelength modes are critically
slowed down and relax to equilibrium on very long 
time scales. Therefore a TDGL description  leads to the conclusion
that if the cooling rate is finite, 
the long-wavelength modes will fall out of LTE and become quenched. As
the temperature falls below the critical, 
these modes will become unstable and will grow exponentially. 

\subsection{Linear instability analysis:}

Let us consider now the situation for $T<<T_c$ and neglect the thermal
noise. The early time evolution after the 
quench is obtained by  linearizing the TDGL equation around a homogeneous
mean field  solution $M_o(t)$. Writing
\begin{equation}
M(\vec r,t) = M_o(t) + \frac{1}{\sqrt{\Omega}} \sum_{\vec k \neq 0}
m_k(t) \; e^{i \vec k\cdot \vec r}
\end{equation} 
where $\Omega$ is the volume of the system, and considering only the 
linear term in the fluctuations $m_k(t)$ the linearized dynamics is the following: 
{\bf COP:} for $M_o(t)$ the conservation gives
$$
\frac{d M_o(t)}{dt} =0
$$
since $M_o$ is the volume integral of the order parameter [see
eq.(\ref{equcop})]   and for the fluctuations we obtain
\begin{equation}
\frac{d m_k(t)}{dt} =   \omega(k) \; m_k(t) ~~;  ~~ \omega(k)  =
-\Gamma_0 \; k^2\left[k^2 + \left. 
\frac{\partial^2 V[M]}{\partial M^2}\right|_{M_o} \right] \label{freqcop}
\end{equation}
In the spinodal region 
$\left.\frac{\partial^2 V[M]}{\partial M^2}\right|_{M_o} <0$
there is a band of  unstable wave vectors $k^2 < 
\left| \left. \frac{\partial^2 V[M]}{\partial M^2}\right|_{M_o}\right|
$ for which 
the frequencies are positive and the fluctuations away from the mean
field grow exponentially. 

{\bf NCOP:} separate the $\vec k\neq 0$ from the $\vec k=0$ in the
linearized equation of motion:
\begin{eqnarray}
\frac{d M_o(t)}{dt} & = &  -\Gamma_0 \left.\frac{d V[M]}{dM}\right|_{M_o}
\label{zeromodelang} \\
\frac{d m_k(t)}{dt} & = & -\Gamma_0 \left[\frac{\delta F[M]}{\delta M}
\right]_{M_o(t)} m_k(t)~  = -\Gamma_0 \left[k^2 + \left.
\frac{\partial^2 V[M]}{\partial M^2}\right|_{M_o} \right] \label{kmodeslang}
\end{eqnarray}
whereas the first equation (\ref{zeromodelang}) determines that $
M_o(t) $
rolls down the potential hill towards the equilibrium solution,
the second equation also displays the linear instabilities for the
same band of wave vectors as in the COP in the spinodal region $ |M_o(t)|\leq
M_s(T) $ [see eq. (\ref{spinoregion})] for which the fluctuations grow
exponentially in time. Thus in the linearized approximation for 
both NCOP and the COP the spinodal 
instabilities are manifest as exponentially growing fluctuations. These
instabilities are the hallmark of the process of phase separation and 
are the early time indications of the formation and growth of correlated
regions which will be understood in an exactly solvable example below.

\subsection{The scaling hypothesis: dynamical length scales for ordering}
The process of ordering is described by the system developing ordered regions
or domains that are separated by walls or other type of defects. 
The experimental probe to study the domain structure and the emergence of
long range correlations is the equal time pair correlation function
\begin{equation}
C(\vec r,t) = \langle M(\vec r,t) M (\vec 0,t) \rangle
\end{equation} 
where $\langle \cdots \rangle$ stands for the statistical ensemble average
in the initial state (or average over the noise in the initial state before
the quench) and will become clear(er) below.  It is convenient to expand
the order parameter in Fourier components
$$
M(\vec r,t) = \frac{1}{\sqrt{\Omega}} \sum_{\vec k} m_k(t)  \; 
e^{i\vec k\cdot\vec x}
$$
and to consider the spatial Fourier transform of the pair correlation function
\begin{equation}
S(\vec k,t) = \langle m_{\vec k}(t) m_{-\vec k}(t) \rangle \label{strucfac}
\end{equation}
known as the {\bf structure factor} or power spectrum which is experimentally
measured by neutron (in ferromagnets) or light scattering (in binary fluids)\cite{goldburg}.
The scaling hypothesis introduces a dynamical length scale $L(t)$ that
describes the typical scale of a correlated region and proposes that
\begin{equation}
C(\vec r,t) = f\left(\frac{|\vec r|}{L(t)}\right) \Rightarrow S(\vec k,t) =
L^d(t) \;  g(kL(t)) \label{scaling}
\end{equation}
\noindent where $d$ is the spatial dimensionality and 
$f$ and $g$ are scaling functions. Ultimately scaling is confirmed by
experiments and numerical simulations and theoretically it emerges from
a renormalization group approach to dynamical critical phenomena which
provides a calculational framework to extract the scaling functions and
the deviations from scaling behavior\cite{bray}. This scaling hypothesis describes
the process of phase ordering as the formation of ordered `domains' or
correlated regions of typical spatial size $L(t)$. For NCOP typical growth
laws are $L(t) \approx t^{1/2}$ (with some systems showing weak logarithmic
corrections) and $L(t) \approx t^{1/3}$ for scalar and $\approx t^{1/4}$
for vector order parameter in the COP case\cite{bray,mazenko,marco}. 

\subsection{An exactly solvable (and relevant) example: the Large $ N $ limit}

We consider the case where the order parameter has $N$-components and transforms
as a vector under rotations in an N-dimensional Euclidean space, i.e.
$\vec M(\vec r,t)= (M_1(\vec r,t), M_2(\vec r,t), \cdots, M_N(\vec r,t))$.
For $N=1$ an example is the Ising model, for $N=2$ superfluids or 
superconductors (where the components are the real and imaginary part of
the condensate fraction or the complex gap respectively), $N=3$ is the spin
one Heisenberg antiferromagnet, etc. For $N=1$ the  topological defects
 are domain walls (topological in one spatial dimension), for $N=2$ they
are vortices in $d=2$ and vortex lines in $d=3$, for $N=d=3$ the topological
defects are monopoles or skyrmions which are possible excitations in
Quantum Hall systems and also appear in nematic liquid
crystals\cite{kibble2}. For  
$N\rightarrow \infty$ and fixed $d$ no topological defects exist. However
the exact solution of the large $ N $ model gives insight and is in fairly
good agreement with growth laws for fixed $ N $ systems which had been studied
experimentally and numerically\cite{bray,marco}. In quantum field
theory the non-equilibrium dynamics of  
phase transitions  has been studied  in Minkowsky and cosmological
space-times\cite{FRW,nuestros,noscorre,losalamos,new}. In cosmological
space-times  it has
been implemented to study the collapse of texture-like
configurations\cite{durrer,turok,filipe} (see 
later). The large $ N $ limit is an exactly solvable limit that serves
as a testing 
ground for establishing the fundamental concepts and that can be systematically
improved in a consistent $1/N$ expansion. It provides a consistent
formulation which is {\em non-perturbative}, 
renormalizable and numerically implementable and has recently been
invoked in novel studies of non-equilibrium dynamics in quantum spin glasses
and disordered systems\cite{cugliandolo}.  

 The exact solution for the dynamics in the large $ N $ limit, being
available both in 
the condensed matter TDGL description of phase ordering kinetics and
in Quantum Field 
Theory in Minkowsky and Cosmological space times, allow us to compare
{\em directly} the physics of phase ordering in these situations. Thus
we begin by implementing this 
scheme in the NCOP case for the TDGL description. 
 
What is the $ \langle \cdots \rangle $ in the equations of the previous
section?: consider that 
{\em before} the quench the system in in equilibrium in the {\em disordered}
phase at $ T>>T_c $ and with a very short correlation length ($ \xi(T)
\approx 1/T $). The  
ensemble average in this initial state is therefore 
\begin{eqnarray}
\langle M^i(\vec r,0) M^j({\vec r}',0) \rangle & = & \Delta  \;
\delta^{ij} \delta^3(\vec r-{\vec r}') \nonumber \\ 
\langle M^i(\vec r,0)  \rangle & = & 0 \label{inicor}
\end{eqnarray} 
\noindent where $\Delta$ specifies the initial correlation. Now
consider a critical quench where the system 
is rapidly cooled through the phase transition to almost zero
temperature but in the {\em absence} of explicit symmetry 
breaking fields (for example a magnetic field). The average of the
order parameter will remain zero through the 
process of spinodal decomposition and phase ordering. During the
initial stages, linear instabilities will grow 
exponentially with $m^i_k(t) \approx m^i_k(0) \; e^{\omega(k) t}~~; ~~
\omega(k)= k^2-r(0)$ for $k^2 < r(0)$ and 
at early times
\begin{equation}
\langle m^i_{\vec k}(t) m^j_{-\vec k}(t) \rangle \approx \Delta \;
e^{2\omega(k)t} 
\end{equation}  
hence fluctuations begin to grow exponentially and eventually will
sample the broken symmetry states and the exponential 
growth must shut-off.
The large $ N $ limit is implemented by writing the potential term in
the free energy as 
\begin{equation}
V[\vec M] = -\frac{r(T)}{2} \; \vec M^2 + \frac{\lambda}{4N} (\vec M^2)^2
~; ~~ \vec M^2= \vec M \cdot \vec M 
\label{largenpot}
\end{equation}
where $\lambda$ is kept finite in the large $ N $ limit. We will focus
on the NCOP case with a quench to zero temperature and rescale the
order parameter, time and space 
as 
\begin{equation}
\vec{M} = \sqrt{\frac{r(0)}{\lambda}} \; \vec \eta ~~; ~~ r(0) \;\Gamma_0 \; t
= \tau ~~; ~~ \sqrt{r(0)} \;\vec x = z 
\label{rescalevars}
\end{equation}
after which the evolution equation for the NCOP case becomes
\begin{equation}
\frac{\partial \vec \eta}{\partial \tau} = \nabla^2 \vec \eta + \left(1-\frac{{\vec{\eta}}^2}{N}\right)~\vec \eta
\end{equation}
where derivatives are now with respect to the rescaled variables. The large $ N $ limit is solved by implementing
a Hartree-like factorization\cite{bray}
\begin{equation}
\vec \eta^2 \rightarrow \langle \vec \eta^2 \rangle =  N  \langle \eta^2_i \rangle ~~ \mbox{no sum over i}
\label{largeNfactori}
\end{equation}
Then for each component the NCOP equation becomes
\begin{eqnarray}
\frac{\partial  \eta_i}{\partial \tau} & = & \left[\nabla^2
+M^2(t)\right] \eta_i \label{effeqn} \\ 
M^2(t) & = & 1-\langle \eta^2_i \rangle  \label{selfconsncop}
\end{eqnarray}
the eq.(\ref{selfconsncop}) is a {\em self-consistent} condition that
must be solved simultaneously with the 
equation of motion for the components. Thus the large $ N $
approximation linearizes the problem at the expense of 
a self-consistent condition. The solution for each component is obviously
\begin{equation}
\eta_{i,\vec k}(\tau) = \eta_{i,\vec k}(0) \;e^{-k^2 \tau + b(\tau)}
~; ~~ b(\tau) = \int^{\tau}_0 M^2(\tau')d\tau' 
\end{equation}
Consider for a moment that the $\vec k=0$ mode is slightly displaced
at the initial time, then it will roll 
down the potential hill to a final equilibrium position for which
$ M^2(\infty) \;\eta_i(\infty) =0 $ (so the 
time derivative vanishes in equilibrium). If $ \eta_i(\infty)\neq 0 $ is
a broken symmetry minimum of the free energy, 
then $ M^2(\tau) \rightarrow  0$ when $ \tau \rightarrow \infty $. This is
the statement of Goldstone's theorem that 
guarantees that the perpendicular fluctuations are soft modes. This
asymptotic limit allows the solution of the 
self-consistent condition  
\begin{equation}
M^2(\tau) = 1-\langle \eta^2_i(\tau) \rangle = 1- {\Delta} \; e^{2
b(\tau)} \int \frac{d^d k}{(2\pi)^d} \;  e^{-k^2 \tau} = 
1- {\Delta} \;  e^{2b(\tau)} \; (8\pi t)^{-\frac{d}{2}}
\end{equation}
The vanishing of the right hand side in the asymptotic time regime
leads to the self-consistent solution 
\begin{equation}
b(\tau) \rightarrow \frac{d}{4} \ln\left[\frac{\tau}{\tau_0}\right] ~~ \Rightarrow M^2(\tau) \rightarrow \frac{d}{4\tau}
\end{equation}
where $\tau_0$ is a constant related to $\Delta$. This self-consistent solution results in the following
 asymptotic behavior
\begin{equation}
\eta_{i,\vec k}(\tau) \rightarrow \eta_{i,\vec k}(0)\left(\frac{\tau}{\tau_0}\right)^{\frac{d}{4}} e^{-k^2\tau}
\end{equation}
Introducing the {\em dynamical length scale} $L(\tau)= \tau^{\frac{1}{2}}$ it is straightforward to find
the structure factor and the pair correlation function
\begin{eqnarray}
&&S(\vec k,t) \propto L^d(t) \; e^{-2(kL(t))^2} \label{struclargen} \\
&&C(\vec r, t) \propto e^{-\frac{r^2}{8L^2(t)}} ~~;~~ L(t) =
t^{\frac{1}{2}} \label{paircorrlargen} 
\end{eqnarray}
This behavior {\em should not} be interpreted as diffusion, because of the
$L^d(t)$ in eqn. (\ref{struclargen}) which is a result of the self-consistent condition.

\underline{\bf Important Features:}
\begin{itemize}
\item{The `effective squared mass' $M^2(t)\stackrel{t\rightarrow
\infty}{\rightarrow} 0$: asymptotically there 
are massless  excitations identified as Goldstone bosons.}

\item{Since $M^2(t)\rightarrow 0$ asymptotically, the self-consistent
condition results in that $\langle \vec M^2 \rangle 
\rightarrow N r(0)/\sqrt{\lambda}$, i.e. the fluctuations sample the
broken symmetry states, which are equilibrium 
minima of the free energy. These fluctuations begin to grow
exponentially at early times due to spinodal instabilities.}  

\item{A dynamical correlation length emerges $L(t)= t^{1/2}$ which
determines the size of the correlated regions or 
`domains'. A scaling solution emerges asymptotically with the
natural scale determined by the size of the ordered 
regions. These regions grow with this law until they become
macroscopically large.  
Although this a result obtained in the large $ N $ limit, similar
growth laws had been found for NCOP both 
analytically and numerically for $N=1$ etc.\cite{bray}} 

\item{{\bf Coarsening:} The expression for the structure factor (\ref{struclargen}) shows that at large times
only the very small wavevectors contribute to $S(\vec k,t)$, however the self-consistency condition forces the
$\int k^{d-1} \; dk \; S(k,t) \rightarrow \mbox{constant}$ thus
asymptotically the structure factor is 
peaked at wavevectors $k \approx L^{-1}(t)$ with an amplitude $L^d(t)$ thus becoming a {\em delta
function} $S(\vec k,t) \stackrel{t\rightarrow \infty}{\rightarrow} \delta^d(\vec k)$. The position of the peak
in $S(\vec k,t)$ moving towards longer wavelength is the phenomenon of coarsening and is observed via light
scattering. At long times a zero momentum condensate is formed\cite{marco} and a Bragg peak develops at zero momentum, this
condensate however grows as a power of time and only becomes macroscopic at asymptotically large times. Coarsening
is one of the experimental hallmarks of the process of phase ordering, revealed for
example in light scattering\cite{goldburg} and is found numerically in many systems\cite{bray}. Thus
the large $ N $ limit, although not being able to describe topological defects offers a very good description of
the ordering dynamics.} 
\end{itemize}

\section{Phase ordering in Quantum Field Theory I: Minkowski space-time}
\subsection{A quench in Q.F.T.}
The dynamics is completely determined by the microscopic field theoretical
Hamiltonian. For a simple scalar theory
the Hamiltonian operator is given by
\begin{equation}
\hat{H} = \int d^3x \left\{ \frac{1}{2}{\Pi}^2(\vec x,t)+
\frac{1}{2}[\vec{\nabla}\Phi(\vec x,t)]^2 + V[\Phi(\vec x,t)]\right\}
\label{qftham} 
\end{equation} 
where $\Phi$ is the quantum mechanical field and $\Pi$ its canonical
momentum. We want 
to describe a quenched scenario where the initial state of the system
for $t<0$ is the ground state (or density matrix,  
see later) of a Hamiltonian for which the potential is convex for all
values of the field, for example that of an 
harmonic oscillator, in which case the wave function(al) $\Psi[\Phi]$
is a Gaussian centered at the origin. At $t=0$ the 
potential is changed so that for $t>0$ it allows for broken symmetry
states. This can be achieved for example 
by the following form
\begin{eqnarray}
V[\Phi] & = & \frac{1}{2} m^2(t) \Phi^2 + \frac{\lambda}{4}\Phi^4 \label{qftpotential} \\
m^2(t) & = & \left\{ \begin{array}{cc}
+m^2_0>0 & ~\mbox{for} ~ t<0 \\
-m^2_0<0 &  ~\mbox{for} ~t>0
\end{array} 
\right. \label{massat}
\end{eqnarray}
Although in Minkowski space-time this is an {\em ad-hoc} choice of a
time dependent potential that 
mimics the quench\cite{bowick}, we will see in the next section that
in a cosmological setting the mass term naturally 
depends on time through the temperature dependence and that it changes
sign below the critical temperature as the 
Universe cools off. Most of the results obtained in Minkowski
space-time will translate onto analogous results in 
a Friedmann-Robertson-Walker cosmology. Unlike the phenomenological
(but succesful) description of the dynamics in 
condensed matter systems, in a microscopic quantum theory the dynamics
is completely determined by the Schr\"odinger 
equation for the time evolution of the wave function or alternatively
the Liouville equation for the evolution of 
the density matrix in the case of mixed states. We will cast our study
in terms of a density matrix in general, such 
a density matrix could describe pure or mixed  states and  obeys the
quantum Liouville equation 
\begin{equation}
i \frac{\partial \hat{\rho}(t)}{\partial t} =
\left[\hat{H}(t),\hat{\rho}(t)\right] \label{liouville}  
\end{equation} 
{\bf Question:} How does the wave function(al) or the density matrix
evolve after a quench?  
\subsection{A simple quantum mechanical picture:}
In order to gain insight into the above
question, let us consider a simple case of one quantum mechanical
degree of freedom $q$ and the quench is 
described in terms of an harmonic oscillator with a time dependent frequency 
$\omega^2(t)= -\epsilon(t) \; \omega^2_0 ~;~\omega^2_0>0$
with $ \epsilon(t) $ the sign function, so that 
$\omega^2(t<0)>0 ~;~\omega^2(t>0)<0$. Furthermore let us focus 
on the evolution of  a pure state (the density matrix is simple the product
of the wave function and its complex conjugate). Consider that at $t<0$ the
wave function corresponds to the ground state of the (upright) harmonic
oscillator.  For $t>0$ the wave function
obeys
\begin{equation}
i\frac{\partial \Psi[q,t]}{\partial t} = \left[-\frac{1}{2}
\frac{d^2}{dq^2}-\frac{1}{2}\omega^2_0\; q^2 \right]\Psi[q,t] 
\end{equation}  
Since the initial wave function is a gaussian and under time evolution with a quadratic Hamiltonian Gaussians remain
Gaussians, the solution of this Schr\"odinger equation is given by
\begin{eqnarray}
\Psi[q,t] & = & N(t) \;  e^{-\frac{A(t)}{2}q^2} \label{qmwf} \\
 \frac{d ln N(t) }{dt} & = & -\frac{i}{2} A(t)  \label{qmunit} \\
i \frac{dA}{dt} & = & A^2 + \omega^2_0 \label{qmkern} \nonumber
\end{eqnarray}

Separating the real and imaginary parts of $A(t)$ it is straightforward to
find that $ |N(t)|^4/\mbox{Re}[A(t)] $ is constant,  a consequence
of unitary time evolution. Eq.(\ref{qmkern}) can be cast in a more
familiar form by a simple substitution
\begin{equation}
A(t) = -i\frac{\dot{\phi}(t)}{\phi(t)} \Rightarrow \ddot{\phi}(t)-\omega^2_0
 \; \phi(t)=0
\end{equation}
where the equation for $\phi$ was obtained by inserting the above expression
for $A(t)$ in (\ref{qmkern}). The solution is $\phi(t) = a \;  e^{\omega_0~t}+
b \;  e^{-\omega_0~t}$ featuring exponential growth. This is the
quantum mechanical 
analog of the spinodal instabilities described in the previous section. 
The equal time two-point function is given by
\begin{equation}
\langle q^2 \rangle(t) = A^{-1}_R(t) = |\phi(t)|^2 \approx
e^{2\omega_0 \, t}
\end{equation}
The width of the Gaussian state increases in time (while the amplitude
decreases to maintain a constant norm) and the quantum fluctuations grow
exponentially. As the Gaussian wave function spreads out the probability for
finding configurations with large amplitude of the coordinates increases.
These is the quantum mechanical translation of the linear spinodal
instabilities. When the non-linear contributions to the quantum mechanical
potential are included the single particle quantum mechanical
wave function will simply develop two peaks and eventually re-collapse by
focusing near the origin undergoing oscillatory motion between `collapses'
and `revivals'. In the case of a full quantum field theory
there are infinitely many degrees of freedom and the energy is transferred
between many modes. This simple quantum mechanical example paves
the way for understanding in a simple manner the main features of a quench in the
large $ N $ limit in quantum field theory, to which we now turn our attention. 
\subsection{Back to the original question: Large $ N $ in Q.F.T.}
We now consider the large $ N $ limit of a full Q.F.T. in which
\begin{equation}
\vec{\Phi}(\vec x,t)= \left(\Phi_1(\vec x,t), \Phi_2(\vec x,t), \cdots,
\Phi_N(\vec x,t) \right) 
\end{equation}
and similarly for the canonical momenta $\vec{\Pi}$. The Hamiltonian
operator is of the form (\ref{qftham})
with
\begin{equation}
V[\vec \Phi]  =  \frac{1}{2} m^2(t) \;  \vec{\Phi}\cdot\vec{\Phi}
 + \frac{\lambda}{8N}[\vec{\Phi}\cdot\vec{\Phi}]^2 \label{qftpotlargen} 
\end{equation}
with $m^2(t)$ given by (\ref{massat}). Let us focus on the case in which 
the initial state pure and symmetric, i.e. $\langle \Phi \rangle =0$, with $<\cdots>$ being
the expectation value in this initial state. The more complicated case
of a mixed state, described 
by a density matrix is studied in detail
in\cite{nuestros,noscorre,losalamos} and the main features are the
same as those revealed 
by the simpler scenario of a pure state.  The large $ N $ limit is implemented in a similar manner as in the
TDGL example, via a Hartree like factorization
\begin{equation}
(\vec{\Phi}\cdot \vec{\Phi})^2 \rightarrow 2\, \langle \vec{\Phi}\cdot
\vec{\Phi}\rangle \; \vec{\Phi}\cdot \vec{\Phi} 
\label{largenqft}
\end{equation}  
where the expectation value is in the time evolved quantum state (in the Schr\"odinger picture) or in the initial
state of the Heisenberg operators (in the Heisenberg picture). Via this factorization the Hamiltonian becomes quadratic at
the expense of a self-consistent condition as it will be seen below. It is convenient to introduce the spatial Fourier transform of the fields as
\begin{equation}
\vec{\Phi}(\vec x,t) = \frac{1}{\sqrt{\Omega}} \sum_{\vec k}
\vec{\Phi}_{\vec k}(t) \;  e^{i \vec k \cdot \vec x} 
\end{equation}
with $\Omega$ the spatial volume, and a similar expansion for the
canonical momentum $\Pi(\vec x,t)$. The Hamiltonian becomes
\begin{eqnarray}
H & = & \sum_{\vec k} \left\{ \frac{1}{2} \vec{\Pi}_{\vec k}\cdot \vec{\Pi}_{-\vec k}+ 
\frac{1}{2} W^2_k(t)\; \vec{\Phi}_{\vec k}\cdot \vec{\Phi}_{-\vec k}
\right\} \label{hamqftlargen} \\ 
W^2_k(t) & = & m^2(t)+k^2+\frac{\lambda}{2N}\int \frac{d^3k}{(2\pi)^3}\; 
 \langle \vec{\Phi}_{\vec k} \cdot \vec{\Phi}_{-\vec k}  \rangle(t) 
\label{timefreqs} 
\end{eqnarray}
The problem now
has decoupled in a set of infinitely many harmonic oscillators, that
are only coupled through the self-consistent 
condition in the frequencies (\ref{timefreqs}). To induce a quench,
the time dependent mass term has the form proposed in eq. (\ref{massat}).

Just as in the simple quantum mechanical case, we consider the initial
state to be a Gaussian centered at the origin in field space, 
which is the ground state of the (upright) harmonic oscillators for
$t<0$. Since a Gaussian is always a Gaussian under 
time evolution with a quadratic Hamiltonian, we propose the wave
function(al) that describes the (pure) quantum mechanical 
state to be given by
\begin{equation}
\Psi[\vec{\Phi},t] = \Pi_k\left\{ N_k(t) \; e^{-\frac{A_k(t)}{2}
\vec{\Phi}_{\vec k}\cdot \vec{\Phi}_{-\vec k}} \right\} 
~;~~ A_k(t=0)= W_k(t<0) \label{wavefunc}
\end{equation}
Time evolution of this wavefunction(al) is determined by the
Schr\"odinger equation: in the Schr\"odinger representation 
the canonical momentum becomes a differential (functional) operator, 
$\vec{\Pi}_{\vec k} \rightarrow -i\delta/\delta \vec{\Phi}_{-\vec k}$
and the Schr\"odinger equation becomes a functional 
differential equation. Comparing the powers of $\Phi_{\vec k}$ in this
differential equation, one obtains the following evolution 
equations for $N_k(t)$ and $A_k(t)$ 
\begin{eqnarray}
\frac{d}{dt}\ln N_k(t) & = & -\frac{i}{2} A_k(t) 
\label{normqft}\\
i \frac{dA_k(t)}{dt} & = & A^2_k(t)- W^2_k(t) \label{kernqft} 
\end{eqnarray}
As in the single particle case, the constancy of
$ |N_k(t)|^4/\mbox{Re}[A_k(t)] $ is a consequence of unitary time  
evolution. The non-linear equation for the kernel $ A_k(t) $ can be
simplified just as in the single particle case by writing
\begin{equation}
A_k(t) = -i \frac{\dot{\phi}_k(t)}{\phi_k(t)} \Rightarrow 
\ddot{\phi}_k(t)+W^2_k(t)\;  \phi_k(t)=0 \label{modesqft}
\end{equation}
and taking the expectation value of $\Phi^2 $ in this state we obtain
\begin{equation}
\langle \vec{\Phi}_{\vec k} \cdot \vec{\Phi}_{-\vec k}  \rangle(t)= N\;
|\phi_k(t)|^2  
\end{equation}
Hence we find a self-consistent condition much like the one obtained
 in the large $ N $ limit for TDGL. The equations for the mode
functions and the self-consistent condition for $t>0$ are therefore given
by 
\begin{eqnarray}
&& \ddot{\phi}_k(t)+[k^2+M^2(t)]\; \phi_k(t) =  0 \label{eqnqftlargen} \\
&&M^2(t) =  -m^2_0 + \frac{\lambda}{2} \int \frac{d^3k}{(2\pi)^3} |\phi_k(t)|^2
\label{selfconsqft} 
\end{eqnarray}
where the integral in the self-consistent term in (\ref{selfconsqft}) is 
simply $\langle \Phi^2_i \rangle$.
There are two fundamental {\em differences} between the quantum dynamics
determined by the equations of motion and the classical dissipative
dynamics of the TDGL phenomenological description given in sec. II:
\begin{itemize}
\item{The equations of motion and the self-consistency condition equations
(\ref{eqnqftlargen})-(\ref{selfconsqft}) lead immediately to the conservation
of energy\cite{FRW,nuestros}.}
\item{The evolution equations are {\em time reversal invariant}.}
\end{itemize}
These properties must be contrasted to the purely dissipative evolution
dictated by the TDGL equations as is clear from eq. (\ref{diffu}).
 Consider a very weakly coupled theory
$\lambda <<1$ and very early times, then the self-consistent term can
be neglected and we see that for $k^2 <m^2_0$ the modes grow exponentially.
This instability again is the manifestation of spinodal 
growth\cite{erickwu,boyvega,boylee,nuestros,noscorre}. Since the
mode functions grow exponentially, fairly soon, at a time scale $t_s \approx
m^{-1}_0 \ln(1/\lambda)$ the self-consistent term begins to cancel the
negative mass squared and  $M^2(t)$ becomes
smaller. We find numerically that this effective mass vanishes asymptotically,
as shown in Fig. 1. 




\subsection{Emergence of condensates and classicality:}
The physical mechanism here is similar to that in the classical TDGL,
but  in terms of quantum  
fluctuations. The quantum fluctuations with wave vectors inside the
spinodally unstable band grow 
exponentially, these make the $\langle \Phi^2 \rangle$ self-consistent
field to grow non-perturbatively  
large until when $\langle \Phi^2 \rangle \approx m^2_0/\lambda$ when
the self-consistent (mean) field  
begins to be of the same order as $m^2_0$ (the tree level mass
term). At this point the {\em quantum} 
fluctuations become non-perturbatively large and sample field
configurations near the  equilibrium minima of the 
potential. The spinodal instabilities are shutting off since the effective squared mass $M^2(t)$ is vanishing.

When $M^2(t)$ vanishes, the equations for the mode functions become those
of a free massless field, with solutions of the form $\phi_k(t) = A_k
\; e^{ikt}+ B_k \; e^{-ikt}$, whereas for the $k=0$ mode the solution
must be of the form 
$\phi_0(t) = a+bt$ with $a ; b \neq 0$ since the Wronskian of the mode
function and its complex conjugate is 
a constant. This in turn determines that the low $k$ (long wavelength)
behavior of the mode functions is given 
by
\begin{equation}
\phi_k(t) = a \cos kt + b \; \frac{\sin kt}{k} \label{specqft}
\end{equation}

This behavior at long wavelength has a remarkable consequence: at very
long time the power spectrum 
$ |\phi_k(t)|^2 $, which is the equivalent of $S(k,t)$ for TDGL (see
eq. (\ref{strucfac})) is dominated by the small $k$-region, in
particular $ k<<1/t $, with an amplitude that 
grows quadratically with time. Then the structure factor $ S(\vec k,t)
= |\phi_k(t)|^2 $ features a 
peak that moves towards longer wavelengths at longer times and whose
amplitude grows with time in such a  
way that asymptotically $ \int^{\infty}_0 k^2 S(\vec k,t) dk / 2\pi^2 \rightarrow
m^2_0/\lambda $ and the integral is dominated 
by a very small region in $k$ that gets narrower at longer times. This
is the equivalent of {\em coarsening} 
in the TDGL solution in the large $ N $ limit, where the asymptotic
time regime was dominated by the formation of 
a long-wavelength condensate. Fig. 2  shows the power spectrum at
two (large) times displaying clearly the 
phenomenon of coarsening and the formation of a non-perturbative condensate. 





The pair correlation function can now be calculated using this power
spectrum\cite{noscorre} 
\begin{equation}
C(\vec r,t) = \frac{1}{2 \, \pi^2 r} \int_0^{\infty} k \sin kr \; |\phi^2_k(t)|
\; dk \; . 
\end{equation}
At long times and distances the integral is dominated by the very long wavelength modes,
in particular by the term $\propto \sin[kt]/k$ of $\phi_k(t)$, hence the integral can be done analytically and
we find
\begin{equation}
C(\vec r,t) = \frac{A}{r}\; \Theta(2t-r) \label{correqft}
\end{equation}
with $ A $ a constant. This is a remarkable result: the correlation
falls off as $1/r$ inside domains that 
grow at the speed of light.  This correlation function is shown in Fig. 3  at several different (large) times. This
 correlation function is of the {\em scaling form}: introducing the
dynamical length scale $L(t) = t$ it is clear that\cite{noscorre} 
\begin{equation}
C(\vec r,t) \propto L^{-1}(t) f(r/L(t)) ~~; ~~ f(s) =
\frac{\Theta(2-s)}{s} \label{scalingqft} 
\end{equation}

We interpret these `domains' as being a non-perturbative condensate of
Goldstone bosons, 
with a non-perturbatively large number of them $\propto 1/\lambda$,
such that the mean square root fluctuation of the field samples the
(non-perturbative) 
equilibrium minima of the potential. In particular an important
conclusion of this analysis is that the long-wavelength 
modes acquire very large amplitudes, their phases vary slowly as a
function of time (for $k<<1/t$), therefore these 
fluctuations which began their evolution as being quantum mechanical,
now have become {\em classical}.




\subsection{O.K...O.K. but where are the defects?}
At this point our analysis begs this question. To understand the
answer it is convenient to back track the analysis to 
the beginning. The initial quantum state is given by a the
wave-function(al) (\ref{wavefunc}), thus the most 
probable field configurations found in this ensemble are those whose
spatial Fourier transform are given by 
\begin{equation}
|\Phi_k| \propto \frac{1}{\sqrt{W_k(t<0)}} \propto \frac{1}{\sqrt{k^2+m^2_0}}
\end{equation} 
(restoring $\hbar$ would multiply $\Phi_k$ by $\sqrt{\hbar}$). Then typical long-wavelength field configurations
that are represented in the quantum ensemble described by this initial wave-function(al) are of rather small 
amplitude. The initial correlations are also rather short ranged on scales $m^{-1}_0$. Under time evolution  the probability distribution is given by 
\begin{equation}
{\cal P}[\Phi,t] = |\Psi[\Phi,t]|^2 = \Pi_{i=1}^N\Pi_k\left\{|N_k(t)|^2 e^{-\frac{|\Phi^i_k(t)|^2}{|\phi_k(t)|^2}}\right\}
\end{equation} 
At times longer than the regime dominated by the exponential growth of the spinodally unstable modes, the power
spectrum $|\phi^2_k(t)|^2$ obtains the largest support for long wavelengths $k<<m^2_0$ and with amplitudes 
$\approx m^2_0/\lambda$. Therefore field configurations with typical spatial Fourier transform $\phi_k(t)$ are very
likely to be found in the ensemble. These field configurations are primarily made of long-wavelength modes and their
amplitudes are non-perturbatively large, of the order of the amplitude of the fields in the broken symmetry minima. 
A typical such configuration can be written as
\begin{equation}
\Phi^i(\vec x,t)_{typical} \approx \sum_k |\phi_k(t)| \cos[\vec k\cdot \vec x + \delta^i_{\vec k}]
\end{equation}
where the phases $\delta^i_{\vec k}$ are randomly distributed with a
Gaussian probability distribution since the density 
matrix is gaussian in this approximation. We note
that a particular choice of these phases leads to a realization of
a likely configuration in the ensemble that 
{\em breaks translational invariance}. In fact translations can be
absorbed by a change in the phases, thus averaging 
over these random phases restores translational invariance. Since the
quantum state (or density matrix) is translational 
invariant a particular spatial profile for a field configuration
corresponds to a particular representative of 
the ensemble. Combining all of the above results together we can
present the following consistent interpretation of the ordering
process and the formation of coherent non-perturbative structures
during the dynamics of symmetry breaking in the large $ N $
limit\cite{noscorre} : 
\begin{itemize}
\item{The early time evolution occurs via the exponential growth of
spinodally unstable long wavelength modes. This 
unstable growth leads to a rapid growth of fluctuations $\langle
\Phi^2 \rangle(t)$ which in turn increases the self-consistent
contribution and tends to cancel the negative mass squared. The
effective mass of the excitations $-m^2_0+ 
\frac{\lambda}{2N}\langle \Phi^2 \rangle(t) \rightarrow 0$ and the
asymptotic excitations are Goldstone bosons.} 

\item{At times larger than the spinodal time $t_s \approx m^{-1}_0
\ln(1/\lambda)$, the effective mass vanishes and 
the power spectrum or structure factor $S(k,t)=|\phi_k(t)|^2$ displays
the features of coarsening: a peak that moves towards 
longer wavelengths and increases in amplitude, resulting in a
long-wavelength condensate at asymptotically long times.}

\item{For large time a dynamical correlation length emerges $L(t) =t$ and at
 long distances the pair correlation function is of the scaling form
$C(\vec r,t) \propto L^{-1}(t) f(r/L(t))$. The length scale $L(t)$
determines the size of the correlated 
regions and determines that these regions grow at the speed of
light. Inside these regions there is a non-perturbative 
condensate of Goldstone bosons with a typical amplitude of the order
of the value of the homogeneous field at the equilibrium broken
symmetry minima.}

\end{itemize}
The similarity between these results and those of the more
phenomenological TDGL description in condensed matter systems 
is rather striking. The features that are determined by the structure
of the quantum field theory are\cite{noscorre}: i) the scaling
variable $ s=r/t $ with 
equal powers of distance and time is  a consequence of the Lorentz
invariance of the underlying  
theory, ii) the fact that the pair correlation function vanishes for
$ r>2t $ is manifestly a consequence of causality.  
An analysis of the correlations and defect density during the spinodal
time scale has been performed in\cite{rivers} 
and related recent studies had been performed in\cite{beilok}.

\section{Phase ordering in Quantum Field Theory II: FRW Cosmology}
\subsection{Cosmology 101 (the basics):}
On large scales $> 100 ~\mbox{Mpc}$ the Universe appears to be
homogeneous and isotropic as revealed by the isotropy 
and homogeneity of the cosmic microwave background and some of the
recent large scale surveys\cite{durrer}. The cosmological principle
leads to a simple form of the metric of space time, the
Friedmann-Robertson-Walker (FRW) metric in terms of a scale factor
that determines the Hubble flow and the 
curvature of spatial sections. Observations seem to favor a flat
Universe for which the space time metric is rather 
simple:
\begin{equation}
ds^2= dt^2 - a^2(t) \; d\vec x^2 \label{FRWmetric}
\end{equation}
the time and spatial variables $ t, \vec x $ in the above metric are
called comoving time and spatial distance 
respectively and have the interpretation of being the time and
distance measured by an observer locally at rest with 
respect to the Hubble flow. At this point we must note that {\em
physical distances} are given  
by $ \vec{l}_{phys}(t)=a(t) \; \vec x $.  An important concept is that of
causal (particle) horizons: events that cannot be connected by 
a light signal are causally disconnected. Since light travels on null
geodesics $ ds^2=0 $ the maximum {\em physical} 
distance that can be reached by a light signal at time $ t $ is given  by 
\begin{equation}
d_H(t) =a(t)\int^t_0 \frac{dt'}{a(t')} \label{horizon} 
\end{equation}
It will prove convenient to change coordinates to {\em conformal time}
by defining a conformal time variable 
\begin{equation} 
\eta = \int^t_0 \frac{dt'}{a(t')} \Rightarrow ds^2 = 
C^2(\eta) \; (d\eta^2 - d\vec x^2) ~~; ~~ C(\eta) = a(t(\eta))\label{conftime}
\end{equation}
in terms of which the causal horizon is simply given by $ d_H(\eta) =
 C(\eta) \;  \eta$ and physical distances as $ \vec{x}_{phys}=C(\eta) \;  \vec
 x $. This metric is of the same form as that of  
 Minkowski space time. 
For energies well below the Planck scale $ M_{Pl} \approx
 10^{19}\mbox{Gev} $ gravitation is well described by {\em classical} 
General Relativity and the Einstein equations:
\begin{equation}
R^{\mu \nu} - \frac{1}{2} \;  g^{\mu \nu} R = \frac{8\pi}{M^2_{Pl}}
\;T^{\mu \nu} \label{einsteineqns} 
\end{equation}
where we have been cavalier and set $c=1$ (as well as $ \hbar=1
$). $ R^{\mu \nu} $ is the Ricci tensor,  $ R $ the Ricci  
scalar and $ T^{\mu \nu} $ the matter field energy momentum
tensor. The above equation is classical but one seeks to 
understand the dynamics of the Early Universe in terms of a {\em
quantum field theory} that describes particle physics, thus the
question: what is exactly 
the energy momentum tensor?, in Einstein's equations it is a classical
object, but in QFT it is an operator. The answer to 
this question is: gravity is classical, fields are quantum mechanical, but 
$ T^{\mu \nu} \rightarrow \langle T^{\mu \nu} \rangle $, i.e. it is the
expectation value of a {\em quantum mechanical operator in a quantum
mechanical state}. This quantum mechanical state, either pure or mixed
is described by a wave-function(al) or a density matrix whose time
evolution is dictated by the quantum equations of motion: the
Schr\"odinger 
equation for the wave functions or the quantum Liouville equation for
a density matrix.  Consistency with the postulate 
of homogeneity and isotropy requires that the expectation value of the
energy momentum tensor must have the fluid form and in the rest frame of the fluid takes the form
$ \langle T^{\mu \nu} \rangle = \mbox{diagonal}(\rho,p,p,p) $ with
$\rho$ the energy density and $p$ the pressure. The time 
and spatial components of Einstein's equations lead to the Friedman equation
\begin{eqnarray}
&&\frac{\dot{a}^2 (t)}{a^2(t)}  =  \frac{8\pi}{3M^2_{Pl}} \rho(t)
\label{hubb} \\ 
&&2\frac{\ddot{a}(t)}{a(t)}+\frac{\dot{a}^2(t)}{a^2(t)}  =
-\frac{8\pi}{M^2_{Pl}} p(t) \label{prss} 
\end{eqnarray}
Combining these two equations one arrives at a simple and intuitive
equation which is reminiscent of the first law 
of thermodynamics:
\begin{equation}
\frac{d}{dt}(\rho a^3(t)) = -p\frac{da^3(t)}{dt} \Rightarrow \dot{\rho}+
3\frac{\dot a}{a}(\rho+p)=0 \label{firstlaw} 
\end{equation}
The alternative form shown on the right hand side of (\ref{firstlaw}) is
the {\em covariant conservation of energy}.
Since the physical volume of space is $ V_0 \;  a^3(t) $ (with $ V_0 $ the
comoving volume) the above equation is recognized as 
$ dU = -p \; dV $ which is the first law of thermodynamics for {\em
adiabatic} processes. To close the set of equations and 
obtain the dynamics we need an equation of state $ p= p(\rho) $: two very
relevant cases are: i) radiation dominated  
(RD) with $ p=\rho/3 $  and matter dominated (MD) $ p = 0 $ (dust)
Universes. In our study we will focus on the RD case.  
The equation of state for RD is that for blackbody radiation for which
the entropy is $ S= C VT^3 $ (with C a constant). 
Since $ V(t)=V_0 \;   a^3(t) $ is the physical volume, the equation
(\ref{firstlaw}) which dictates adiabatic (isoentropic) 
expansion leads to a time dependence of the temperature: $ T(t) =
T_0/a(t) $. Now the cooling is done by the expansion 
of the Universe and a phase transition will occur when the Universe
cools below the critical temperature for a given 
theory. For the GUT transition $ T_c \approx 10^{16} \, \mbox{Gev} \approx
10^{29}K $, for the EW transition 
$ T_c \approx 100 \, \mbox{Gev}\approx 10^{15}K $. Returning now back to
the large $ N $ study of the dynamics of phase transitions, 
we can include the effect of cooling by the expansion of the Universe
by replacing the time dependent mass term $ m^2(t) $ 
in (\ref{qftpotlargen}) by 
\begin{equation}
m^2(t) = m^2_0 \left[\frac{T^2(t)}{T^2_c}-1\right] ~~; ~~ T(t) =
\frac{T_i}{a(t)} \label{masstfrw} 
\end{equation}
This form is consistent with the Landau-Ginzburg description including
the time dependence of the temperature via the 
isentropic expansion of the Universe, but perhaps more importantly it
can be proven in a detailed manner from the  
self-consistent renormalization of the mass in an expanding
Universe\cite{FRW}. Thus the large $ N $ 
limit in a RD FRW cosmology will be studied by using the potential
(\ref{qftpotlargen}) but with the time dependent 
mass given by (\ref{masstfrw}). 
\subsection{Large $ N $ in a RD FRW Cosmology}
The large $ N $ limit is again implemented via the Hartree-like factorization 
(\ref{largenqft}) performing the spatial Fourier transforms of the fields
and their canonical momenta and including the proper scale factors,
the Hamiltonian now becomes\cite{FRW}
 
\begin{eqnarray}
H(t) & = &  \sum_k \left\{ \frac{1}{2a^3(t)} \;\vec{\Pi}_{\vec k}\cdot
\vec{\Pi}_{-\vec k}+ W^2_k(t) \; \vec{\Phi}_{\vec k}\cdot
\vec{\Phi}_{-\vec k} \right\} \label{hamlargenfrw}\\
W^2_k(t) & = & \frac{k^2}{a^2}+m^2(t)+\frac{\lambda}{2N} \langle
\vec{\Phi}_{\vec k}\cdot \vec{\Phi}_{-\vec k}\rangle \label{frwfreqs}
\end{eqnarray}
where now the expectation value is in terms of a {\em density matrix}
$\rho[\Phi(\vec{.}), \tilde{\Phi}(\vec{.});t]$ since we are considering
the case of a thermal ensemble as the initial state. 

We propose the following Gaussian ansatz for the functional density
matrix elements in the {Schr\"{o}dinger} representation\cite{FRW}
\begin{eqnarray}
\rho[\Phi,\tilde{\Phi},t]  =  \prod_{\vec{k}} {\cal{N}}_k(t) \exp\left\{
- \frac{A_k(t)}{2} \;\vec{\Phi}_{\vec k}\cdot \vec{\Phi}_{-\vec k}+
\frac{A^*_k(t)}{2} \;\tilde{\vec{\Phi}}_{\vec k}\cdot 
\tilde{\vec{\Phi}}_{-\vec k}+
B_k(t) \;\vec{\Phi}_{\vec k}\cdot 
\tilde{\vec{\Phi}}_{-\vec k} \right\}
      \label{densitymatrixfrw}
\end{eqnarray}
 This form of the density matrix
is dictated by the hermiticity condition $\rho^{\dagger}[\Phi,\tilde{\Phi},t] =
\rho^*[\tilde{\Phi},\Phi,t]$; as a result of this, $B_k(t)$ is real.
The kernel $B_k(t)$ determines the amount of mixing in the
density matrix, since if $B_k=0$, the density matrix corresponds to a pure
state because it is a wave functional times its complex conjugate. 
The kernels $A_k(0)~~;~~B_k(0)$ are chosen such that the initial density
matrix is thermal with a temperature $T_i > T_c$\cite{FRW}. 
Following
the same steps as in Minkowski space time, the time evolution of this 
density matrix can be found in terms of a set of mode functions $\phi_k(t)$
that obey the following equations of motion and self-consistency condition
\begin{eqnarray}
&&\ddot{\phi}_k(t)+ 3 \; \frac{\dot{a}}{a} \;\dot{\phi}_k(t)+
\left[\frac{k^2}{a^2(t)}+M^2(t)\right]\phi_k(t) = 0 \label{frweqnsofmotion} \\
&&M^2(t) = m^2_0\left[\frac{T^2_i}{T^2_ca^2(t)}-1\right]+\frac{\lambda}{2}
\int \frac{d^3k}{(2\pi)^3} \; |\phi_k(t)|^2 \;
\coth\frac{W_k(0)}{2T_i}\; .
\end{eqnarray}
This equations can be cast in a more familiar form by changing coordinates
to conformal time (see eq. (\ref{conftime})) and (conformally) rescaling the
mode functions $\phi_k(t) = f_k(\eta)/ C(\eta)$ to obtain the following
equations for the conformal time mode functions $f_k(\eta)$ in an RD FRW cosmology
\begin{equation}
f''_k(\eta)+ \left[k^2+ C^2(\eta)M^2(\eta) \right]f_k(\eta) =0 
\label{conftimeqns}
\end{equation}
where primes now refer to derivatives with respect to conformal time. For
RD FRW $C(\eta)= 1+ \eta/2$ (in units of $m^{-1}_0$ which is the only 
dimensionful variable). The above equations of motion has now an analogous
form as those solved in the case of Minkowski space-time. 

As the temperature falls below the critical the effective squared
mass term becomes negative and spinodal instabilities trigger the process
of phase ordering. This results in that the quantum fluctuations
quantified by $\langle \vec \Phi^2 \rangle$ grow 
exponentially. 
These spinodal instabilities make the self-consistent
field  grow at early times and tends to overcome the negative sign of
the squared mass, eventually reaching an asymptotic regime in which the
total effective mass $ M^2(\eta) $ vanishes.


 

 Again this behavior
determines that the fluctuations are sampling the equilibrium broken
symmetry minima of the initial potential, i.e. $\langle \vec \Phi^2
\rangle \rightarrow \frac{2Nm^2_0}{\lambda}$.

Although, just as in Minkowski space-time 
the effective mass vanishes asymptotically,  the non-equilibrium
evolution is rather {\em different}. We find numerically\cite{new}
that asymptotically the effective mass term 
$ C^2(\eta) M^2(\eta) $ vanishes as $ -15/4\eta^2 $.

 Fig. 4 
displays $  C^2(\eta) M^2(\eta) $ as a function of 
conformal time for the case of $ T_i/ T_c = 1.1 $ with 
$ T_c \propto m_0/\sqrt{\lambda} $\cite{FRW,boylee}.

We see that at very
early time the mass is positive, reflecting 
the fact that the initial state is in equilibrium at an initial
temperature larger than the critical. As time evolves 
the temperature is  red-shifted and  cools and at some point the
phase transition occurs, when the mass vanishes 
and becomes negative. 




Figure 5 displays $\frac{\lambda}{2Nm^2_0}\langle \vec{\Phi}^2
\rangle(\eta)$ vs. $\eta$  in units of 
$m^{-1}_0$ for $\frac{T_i}{T_c}=3$, $g=10^{-5}$ for an  R.D. Universe. Clearly at large times 
the non-equilibrium fluctuations probe the broken symmetry states.

This particular asymptotic behavior of the mass determines that 
 the mode functions $ f_k(\eta) $ grow as $ \eta^{5/2} $ for $k < 1 /
 \eta$ and oscillate in the form $ e^{\pm i k \eta}$ for 
$ k > 1 / \eta $. This behavior is confirmed
 numerically\cite{new}. We find both analytically and numerically that 
asymptotically the mode functions are of the {\em scaling form}
\begin{equation}
f_k(\eta) = A \eta^{\frac{5}{2}} \;\frac{J_2(k\eta)}{(k\eta)^2}
\label{scalingFRW} 
\end{equation}
Where $ A $ is a numerical constant and $J_2(x)$ is  a Bessel function.

Figure 6  displays $\eta^{-5}|f_k(\eta)|^2$ as a function of the
scaling variable $k\eta$ revealing the scaling behavior.

It is
remarkable that this is exactly the same scaling solution 
found in the {\em classical} non-linear sigma model in the large $ N $
limit and that describes the collapse of textures\cite{turok},
and also within the context of TDGL equations in the large $ N $ limit
applied to cosmology\cite{filipe}.

The growth of the long-wavelength modes and the oscillatory behavior
of the short wavelength modes again results in that  
the peak of the structure factor $ S(k,\eta) = |f_k(\eta)|^2 $  moves
towards longer 
wavelengths and the maximum amplitude increases.  This is the 
equivalent of coarsening and the onset of a condensate.

Although quantitatively different from Minkowsky space time, the
qualitative features are similar. Asymptotically the non-equilibrium
dynamics results in the formation of a non-perturbative 
condensate of long-wavelength Goldstone bosons. We can now compute the
pair correlation function $C(r,\eta)$ from the mode functions
(\ref{scalingFRW}) and find that it is cutoff by causality at
$r=2\eta$. The correlation function is depicted in Fig. 4  for
two different (conformal) times. 







 The scaling form of the pair
correlation function is
$$
C(r,\eta) = \eta^2 \; \chi(r/\eta)
$$
where $ \chi(x) $ is a hump-shaped function as shown in fig. 7. 

Clearly a {\em dynamical} length scale $L(\eta) = \eta$ emerges as a consequence of causality, much in the
same manner as in Minkowsky space time. The {\em physical} dynamical correlation length is therefore given by
$\xi_{phys}(\eta) = C(\eta) L(\eta) = d_H(t)$,
that is the correlated domains grow again at the speed of light and their size is given by the causal horizon. 
The interpretation of this phenomenon is that within one causal horizon there is one correlated domain, inside which
the mean square root fluctuation
of the field is approximately the value of the equilibrium minima of the tree level potential, this is clearly
consistent with Kibble's original observation\cite{kibble,kibble2}. Inside this domain there
is a non-perturbative condensate of Goldstone bosons\cite{new}.

\section{Conclusions and looking ahead} 

In these lectures we have discussed the multidisciplinary nature of
the problem of phase ordering kinetics and 
non-equilibrium aspects of symmetry breaking. Main ideas from
condensed matter were discussed and presented in 
a simple but hopefully illuminating framework and applied to the rather different realm of phase transitions in quantum field theory as needed to understand cosmology and particle physics. The large $ N $ approximation has provided a bridge that
allows to cross from one field to another and borrow many of the ideas that had been tested both theoretically and
experimentally in condensed matter physics. There are, however, major differences between the condensed matter and
particle physics-cosmology applications that require a very careful treatment of the quantum field theory that cannot
be replaced by simple arguments. The large $ N $ approximation in field theory provides a robust, consistent non-perturbative
framework that allows the study of phase ordering kinetics and dynamics of symmetry breaking in a controlled and consistently implementable framework, it is renormalizable, respects all symmetries and can be improved in a well defined
manner. This scheme extracts cleanly the non-perturbative behavior, the quantum to classical transition and allows to
quantify in a well defined manner the emergence of classical stochastic behavior arising from non-perturbative physics. 
The emergence of scaling and a dynamical correlation length are robust features of the dynamics and the Kibble-Zurek
scenario describes fairly well the general features of the dynamics, albeit the details require careful study, both
analytically and numerically. 

Of course this is just the beginning, we expect a wealth of important phenomena to be revealed beyond the large $N$, such
as the approach to equilibrium, the emergence of other time scales associated with a hydrodynamic description of the
evolution at late times and a more careful understanding of the reheating process and its influence on cosmological
observables. Although within very few years the wealth of
observational data will provide a more clear picture of the
cosmological fluctuations, it is clear that the program that pursues
a fundamental understanding of the underlying physical mechanisms will continue
seeking to provide  a consistent microscopic description of the
dynamics of cosmological phase transitions.  

\section{Acknowledgements:} 
D. B. thanks the organizors of the school, Henri (`Quique') Godfrin
and Yuri Bunkov for a very  
stimulating school and for their warm hospitality and Tom Kibble and
Ana Achucarro for their kind 
invitation and patience. 
 D. B. thanks the N.S.F for
partial support through grant awards: PHY-9605186 and INT-9815064 and LPTHE for warm
hospitality, H. J. de Vega thanks the Dept. of Physics at the Univ. of Pittsburgh for hospitality.
  R. H., is supported by DOE grant DE-FG02-91-ER40682. We
thank NATO for partial support.





\begin{figure}
\centerline{ \epsfig{file=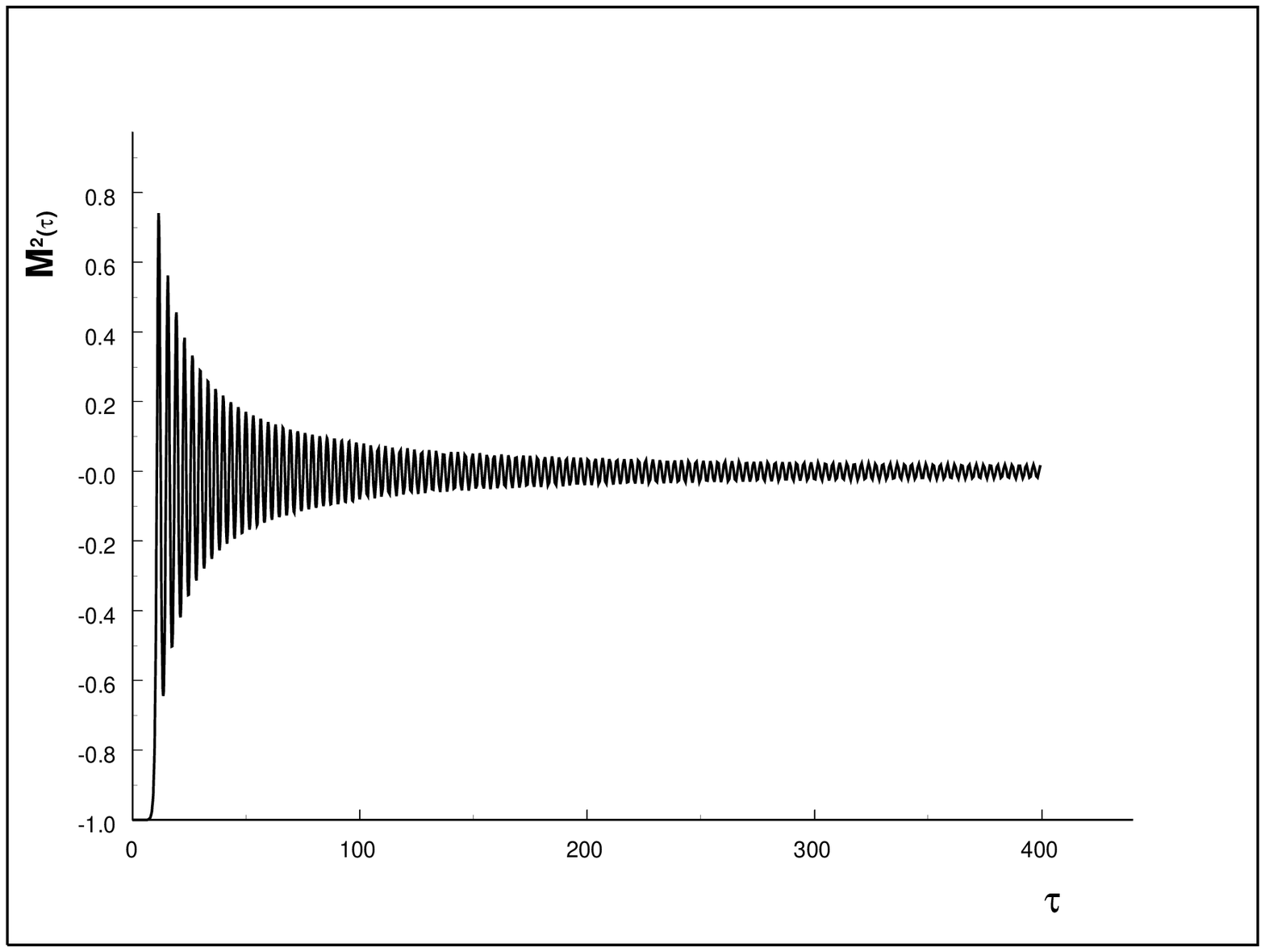,width=7in,height=8in}}
\caption{${\cal M}^2(\tau)$ vs. $\tau$, $g=10^{-7}$ \label{fig1}}
\end{figure}



\begin{figure}
\centerline{ \epsfig{file=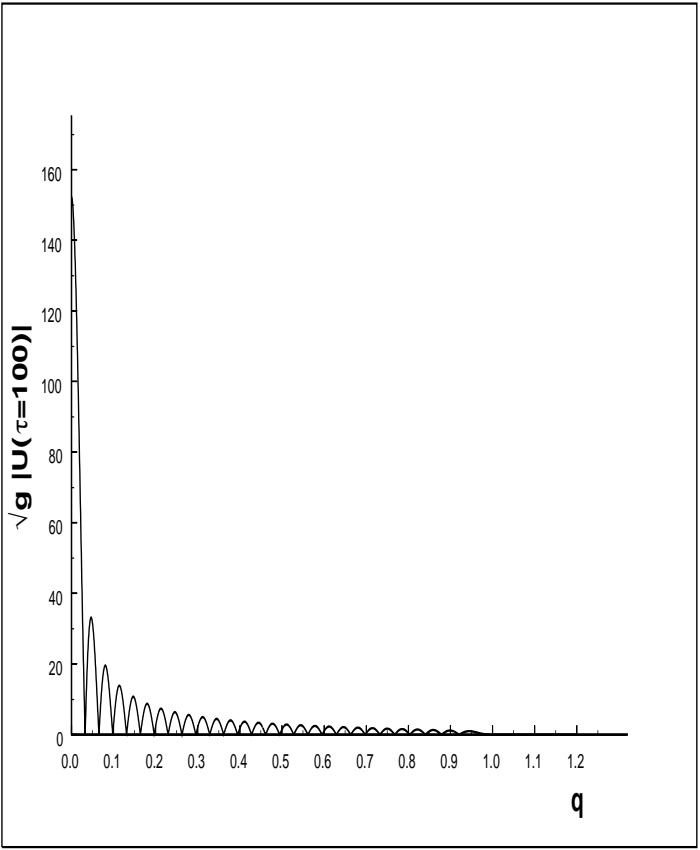,width=4in,height=5in}  \epsfig{file=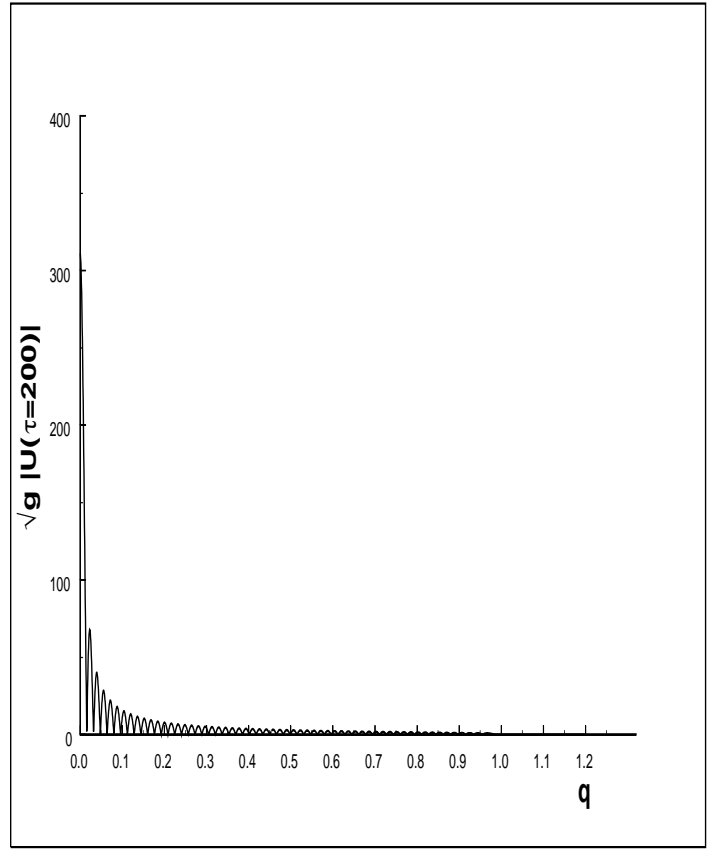,width=4in,height=5in} }
\caption{$g|\phi_k(\tau=100,200)|^2$ vs. $q=k/|m_{0}|$, $g=10^{-7}$ \label{fig2}}
\end{figure}



\begin{figure}
\epsfig{file=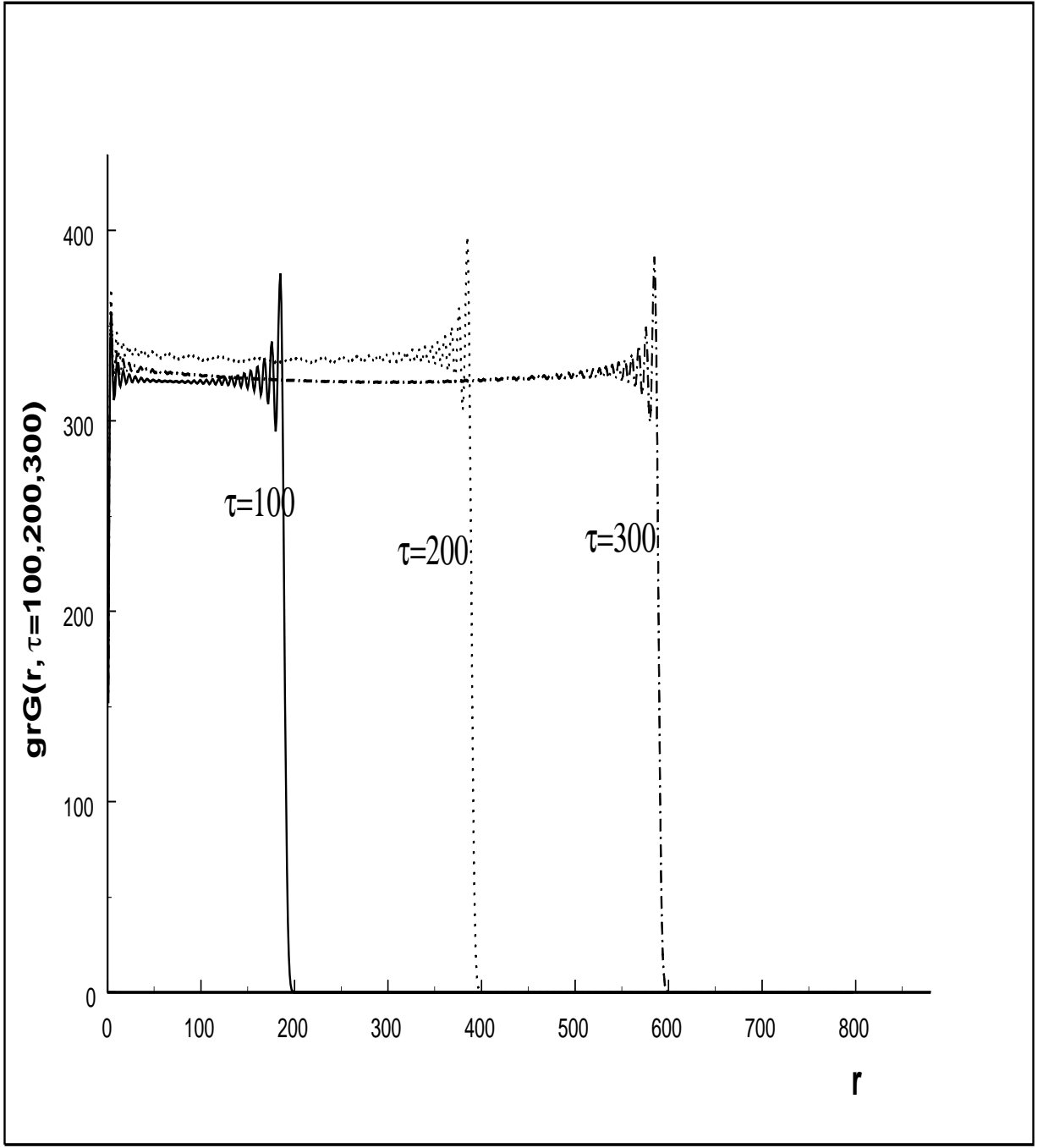,width=7in,height=8in}
\caption{$ g r\, C(r,\tau) $ vs $ r/|m_{0}| $ for $ t/|m_0| = 100, \;
 200, \; 300 $  for $ g = 10^{-7} $.
 \label{fig3}}
\end{figure}



\begin{figure}
{\epsfig{file=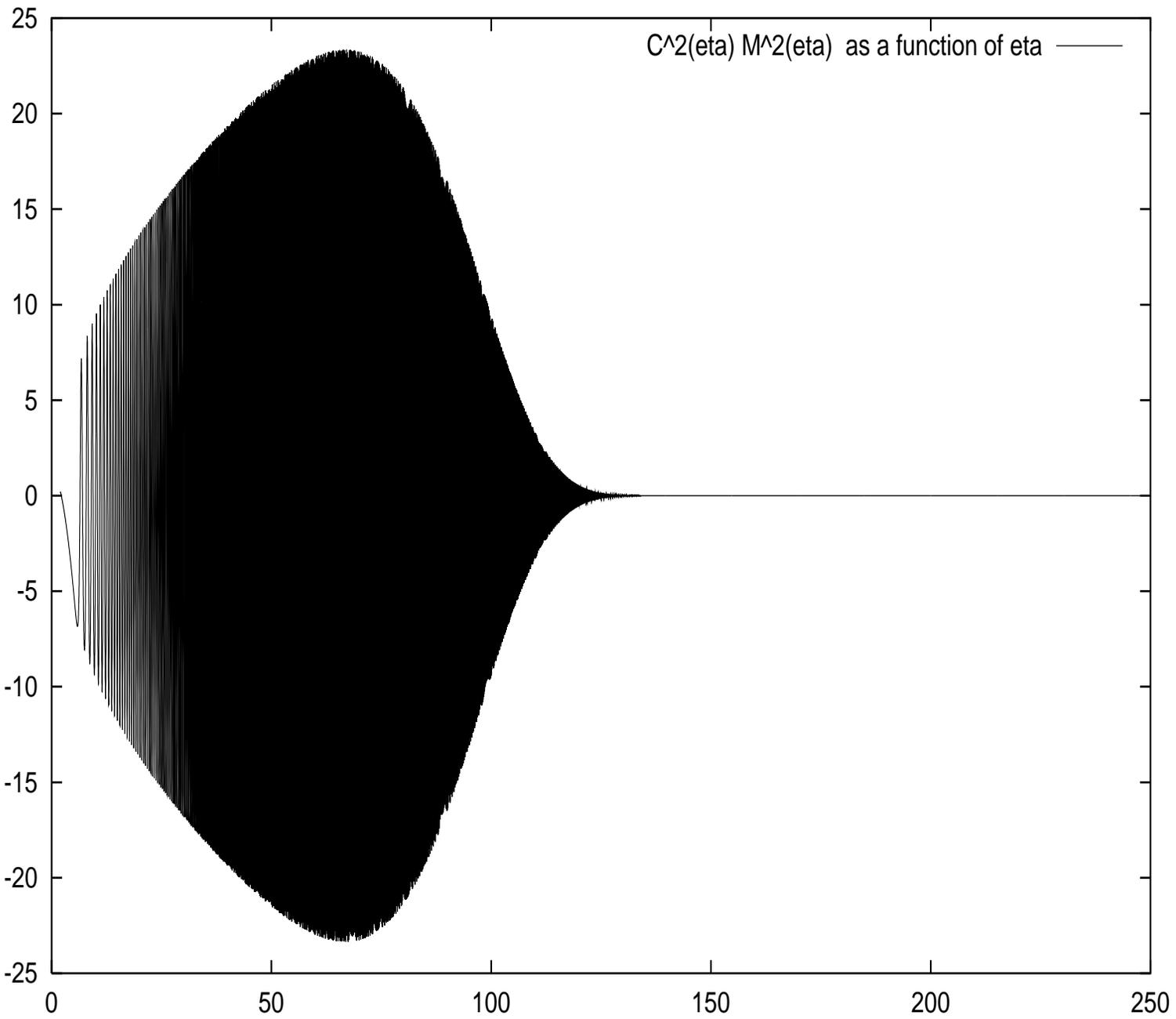,width=7in,height=8in}}
\caption{ $C^2(\eta) M^2(\eta)$ vs. $\eta$(conformal time in units of
$m^{-1}_0$) for $\frac{T_i}{T_c}=3$, $g=10^{-5}$. R.D. Universe.\label{Fig.4} } 
\end{figure}



\begin{figure}
{\epsfig{file=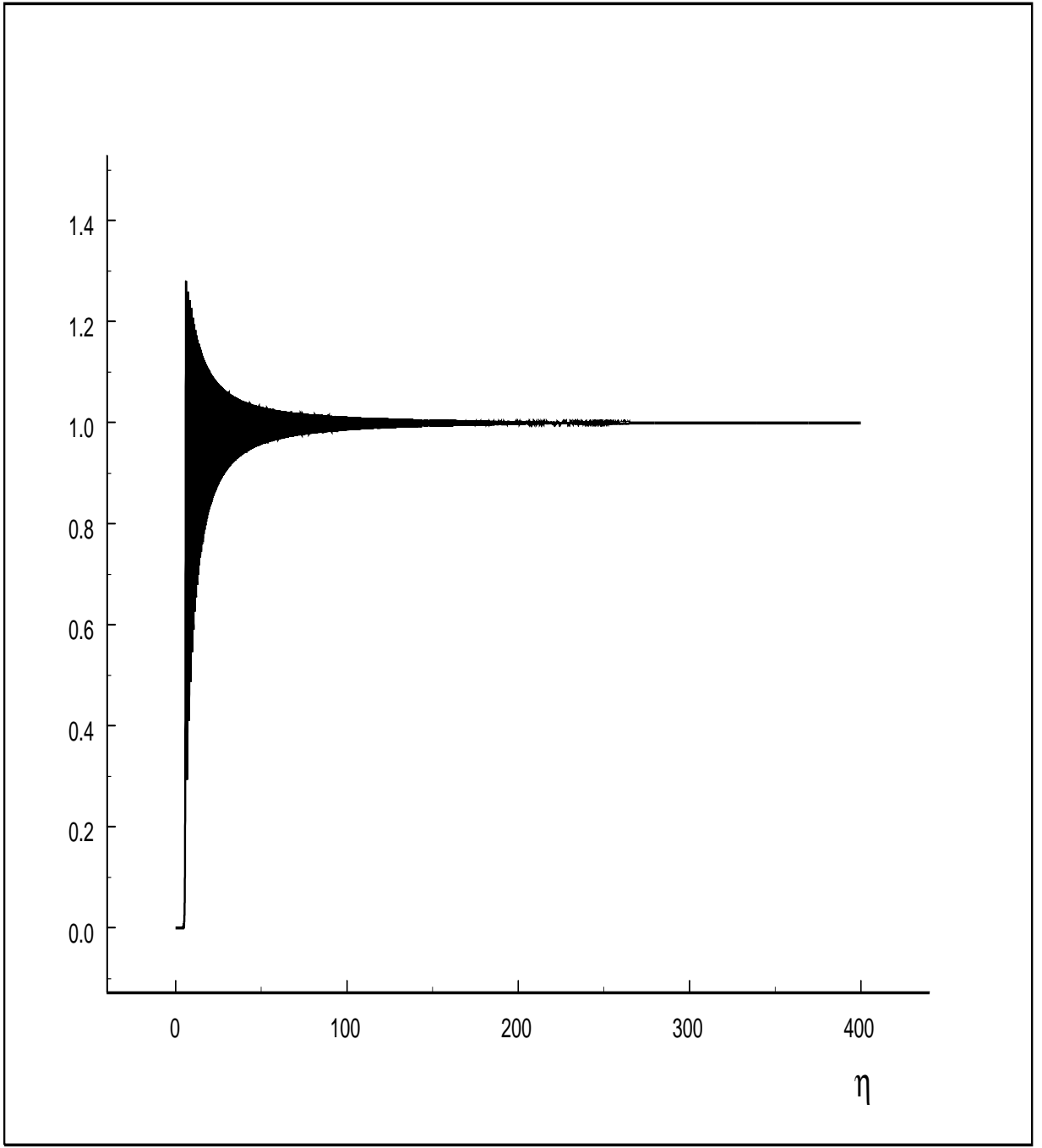,width=7in,height=8in}}
\caption{ $\frac{\lambda}{2Nm^2_0}\langle \vec{\Phi}^2
\rangle(\eta)$ vs. $\eta$ (conformal time in units of 
$m^{-1}_0$) for $\frac{T_i}{T_c}=3$, $g=10^{-5}$.  R.D. Universe.\label{Fig.5} } 
\end{figure}



\begin{figure}
{\epsfig{file=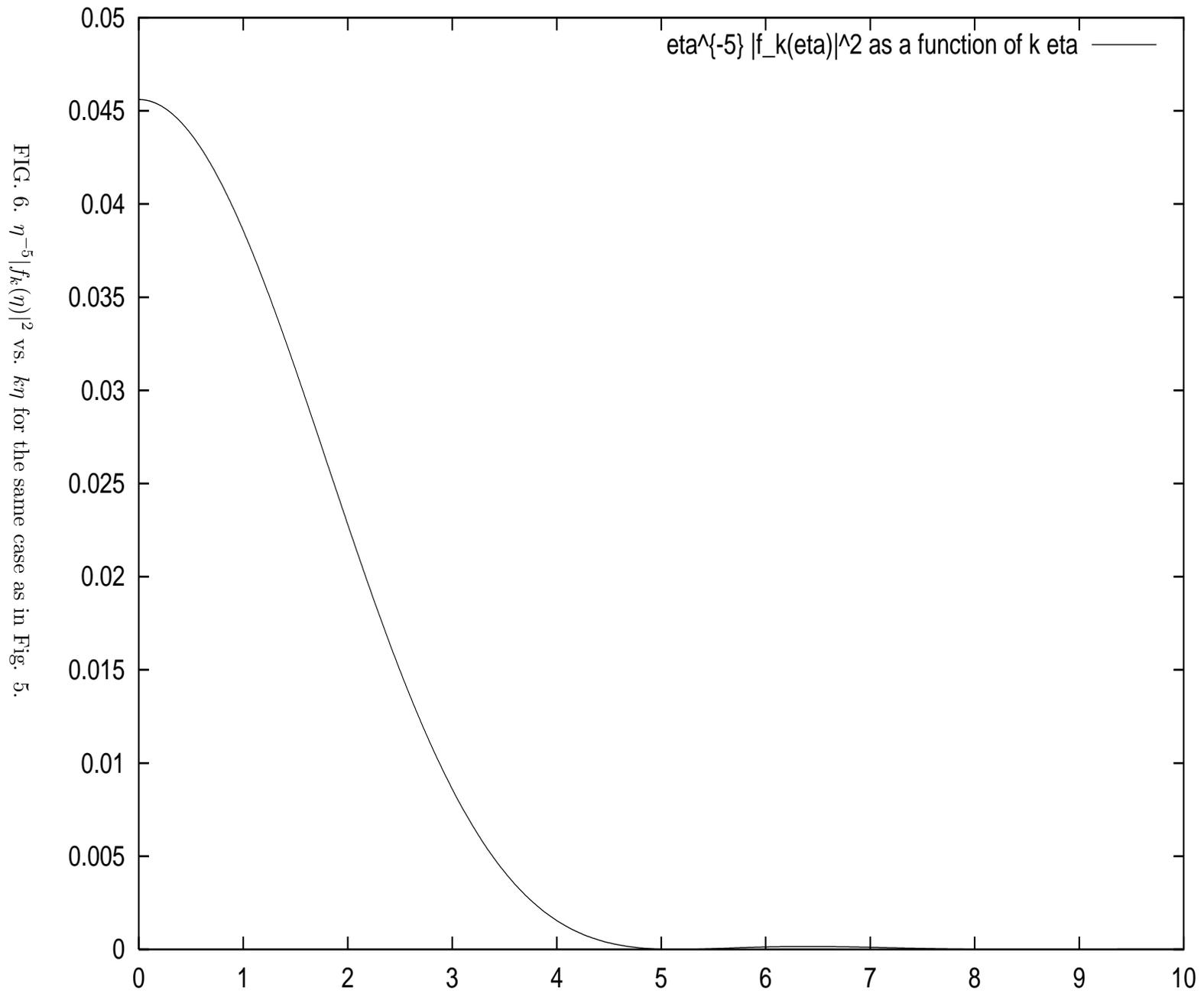,width=7in,height=8in}}
\caption{$\eta^{-5}|f_k(\eta)|^2$ vs. $k\eta$ for the same case as in Fig. 5.\label{Fig.6}}
\end{figure}



\begin{figure}
{\epsfig{file=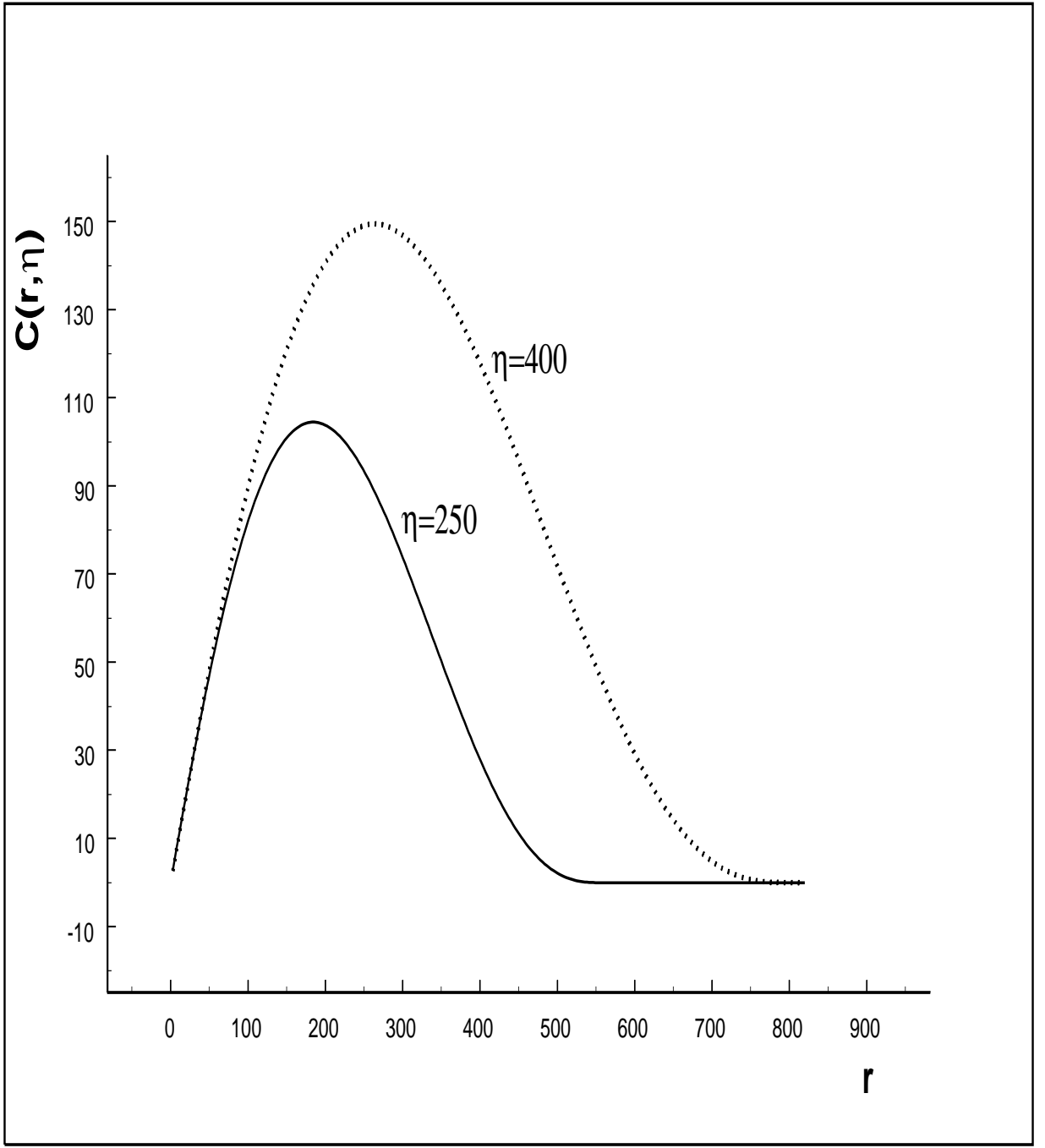,width=7in,height=8in}}
\caption{$C(r,\eta)$ vs. $r$ for $\eta=250,400$ (in units of 
$m^{-1}_0$) for $\frac{T_i}{T_c}=3$, $g=10^{-5}$.  R.D. FRW Universe. \label{Fig.7}}
\end{figure}


\end{document}